\documentclass[10pt, doublecolumn]{IEEEtran}
\usepackage{amsmath}
\usepackage{bbm}
\usepackage{graphicx}
\usepackage{epsfig}
\usepackage{array}
\usepackage{psfrag}
\usepackage{amssymb}
\usepackage{mathdots}
\usepackage{color}
\usepackage{pstricks,pst-node,pst-text,pst-3d,pst-plot}
\usepackage{psfrag}
\usepackage{enumerate}
\usepackage{url,cite}
\usepackage{amsfonts}%
\usepackage{amsthm}
\usepackage{dsfont}%
\usepackage{verbatim}%
\usepackage{setspace}
\usepackage{float}
\usepackage{url}
\usepackage{fancyhdr}
\usepackage{bm}
\usepackage{soul}
\usepackage{lipsum} 

\usepackage{algorithm}
\usepackage{algorithmic}

\usepackage{cleveref}

\newcommand{\figures}{figures}

\newcommand{\Oneik}{\mathds{1}_i(k)}

\newcommand{\Rzk}{R_0(k)}

\newcommand{\Yik}{Y_i(k)}
\newcommand{\bfYk}{{\mathbf Y}(k)}

\newcommand{\gammazik}{\gamma_{0i}(k)}

\newcommand{\gammaijk}{\gamma_{ij}(k)}
\newcommand{\gammaistjk}{\gamma_{\ist j}(k)}

\newcommand{\Gammazk}{\Gamma_0(k)}

\newcommand{\Gammaistk}{\Gamma_{\ist}(k)}

\newcommand{\sN}{\script{N}}

\newcommand{\PRzk}{P_0}

\newcommand{\PRistk}{P_{\ist}\lb k\rb}

\newcommand{\bmuRi}{\overline{R}_i}
\newcommand{\bmuNR}{\overline{\mu}_{\rm II}}

\newcommand{\muRk}{\mu_{\rm I}\lb k\rb}

\newcommand{\muNRk}{\mu_{\rm II}\lb k\rb}

\newcommand{\muopt}{\mu_{\rm II}^{\rm (opt)}}

\newcommand{\Psioptk}{\Psi^{\rm (opt)}(k)}
\newcommand{\ist}{i^*}

\newtheorem{thm}{Theorem}
\newtheorem{lma}{Lemma}
\DeclareMathOperator{\E}{\mathbb{E}}
\newenvironment{proofsketch}{\par{\it Proof Sketch:}}{\qed\par}

\newcommand{\lb}{\left (}
\newcommand{\rb}{\right )}

\newcommand{\script}[1]{{\mathcal {#1}}}

\newcommand{\EE}[1]{\E \left[ #1 \right]}
\newcommand{\EEU}[1]{\E_{\bfU(k)} \left[ #1 \right]}

\newcommand{\bgammaij}{\overline{\gamma}_{ij}}

\newcommand{\bfU}{{\bf U}}

\newcommand{\Ts}{T}

\begin{document}
\title{Offloading Deadline-Constrained Cellular Traffic}

\author{Ahmed Ewaisha, Cihan Tepedelenlio\u{g}lu\\
\small{School of Electrical, Computer, and Energy Engineering, Arizona State University, USA.}\\
\small{Email:\{ewaisha, cihan\}@asu.edu}\\
\date{}
}
\maketitle
\begin{abstract}
In this work we study the problem of hard-deadline constrained data offloading in cellular networks. A single-Base-Station (BS) single-frequency-channel downlink system is studied where users request the same packet from the BS at the beginning of each time slot. Packets have a hard deadline of one time slot. The slot is divided into two phases. Out of those users having high channel gain allowing them to decode the packet in the first phase, one is chosen to rebroadcast it to the remaining users in the second phase. This gives the remaining users a second opportunity to potentially decode this packet before the deadline passes. By this, the BS has offloaded the packet to a ``local network of users'' which eliminates unnecessary BS retransmissions. The problem is modeled as a rate-adaptation and scheduling optimization problem to maximize the duration of this second phase such that each user receives a certain percentage of the packets. We show that the proposed algorithm has a polynomial complexity in the number of users with optimal performance.
\end{abstract}

\section{Introduction}
The 5th generation of wireless communication standards demand more stringent deadlines with higher throughput demands compared to their 4th generation counterparts. Extensive work in the literature has emerged to satisfy these requirements. Offloading the data from the base station (BS) to cellular users was shown to provide promising results to increase the network throughput and users satisfaction \cite{ji2015throughput,le2013instantly,golrezaei2014base,keller2012microcast}. While the algorithms in these works reduce the retransmission traffic significantly, they are suitable for data with no hard deadlines imposed on each packet. Hence, such algorithms are unsuitable for applications such as streaming videos. In addition, the variability of the wireless channel between the BS and the users is ignored in those algorithms. The authors of \cite{7925800} propose a cooperative device-to-device-based communication scheme that improves the cellular network's spectral efficiency. While cooperation is shown to improve the network's performance \cite{7920395,7952823}, offloading the data was out of the scope of their work.

The contributions of this paper is summarized as follows:
\begin{itemize}
	\item Modeling the data offloading problem in the presence of hard deadlines and channel variations.
	\item Presenting a scheduling algorithm with polynomial complexity in the number of users and showing its asymptotic optimality.
\end{itemize}

\section{System Model}
\label{Model}
We assume a single-frequency-channel, time-slotted downlink system with slot duration of $\Ts$ seconds. The system has a single base station (BS) and $N$ users indexed by elements from the set $\sN\in\{1,\cdots,N\}$ while the BS has the index $0$. The users are streaming the same data that is divided into packets that arrive to the BS, each slot, to be broadcast to the users in a timely manner before its hard deadline. We model the channels between the BS and user $i\in\sN$ as a fading channel with power gain $\gammazik\in\mathds{R}_+$ that is known to the BS at the beginning of each slot. The distribution of $\gammazik\in\mathds{R}_+$ can be modeled using the approaches in \cite{8053830} or \cite{fe7af411c5064e66a78e5c890947636d} which present efficient ways of modeling fading channels.

\begin{figure}
\centering
\includegraphics[width=0.8\columnwidth]{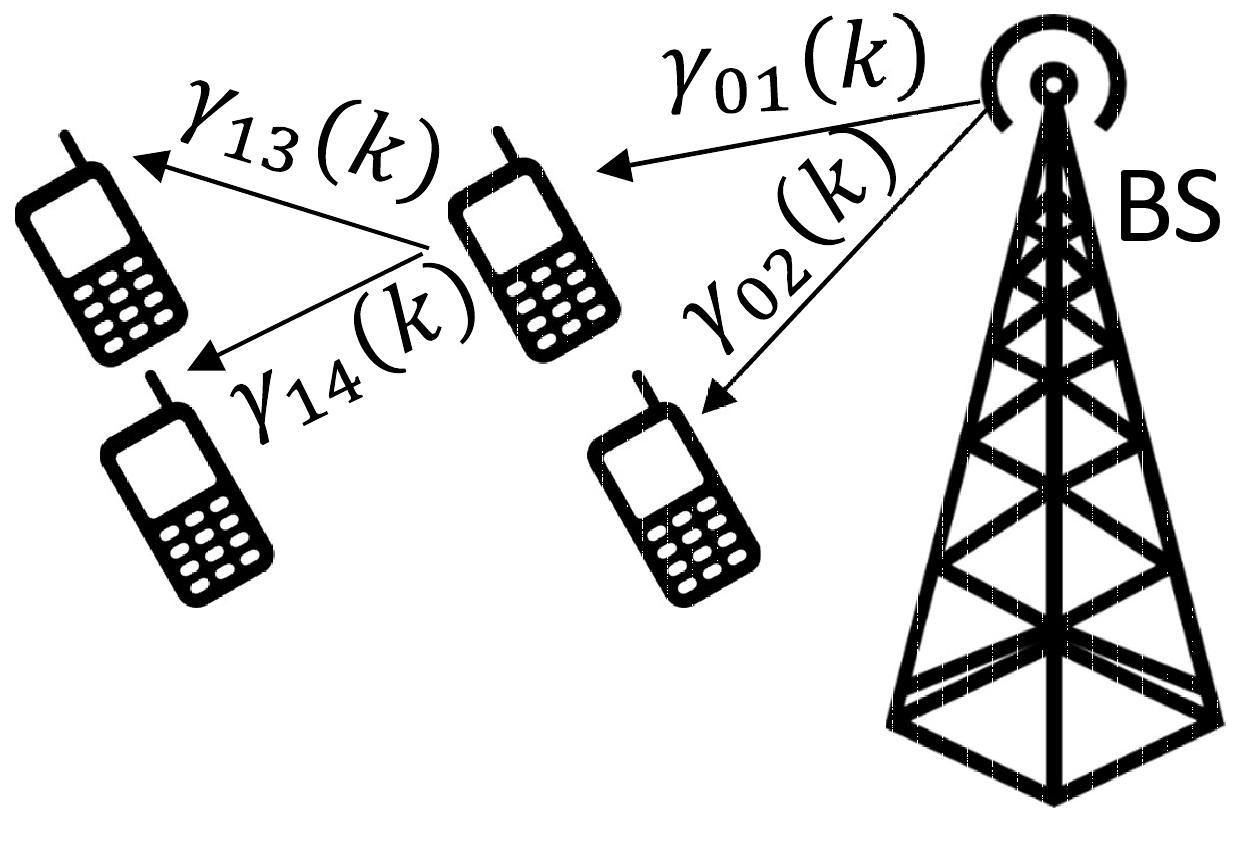}%
\caption{We consider a downlink system where the BS is broadcasting a multicast packet. The packet has a hard deadline and users are allowed to relay to each other as long as no more than one user is transmitting at time.}%
\label{BS}%
\end{figure}

\subsection{Packet Arrival Model}
Let $a(k)$ be the indicator for a packet arrival at the BS at the beginning of slot $k$ and if not received, by some user $i$, by the end of slot $k$ (hard deadline), then this packet is dropped out of the system and does not contribute towards the throughput of that user. Assuming that $\{a(k)\}$ is a Bernoulli process with rate $\lambda$ packets per slot, user $i$ is satisfied if it receives, on average, more than $q_i\%$ of the packets arrived at the BS. We refer to this constraint as the QoS constraint for user $i$.

\subsection{Packet Service Model}
\begin{figure}
\includegraphics[width=1\columnwidth]{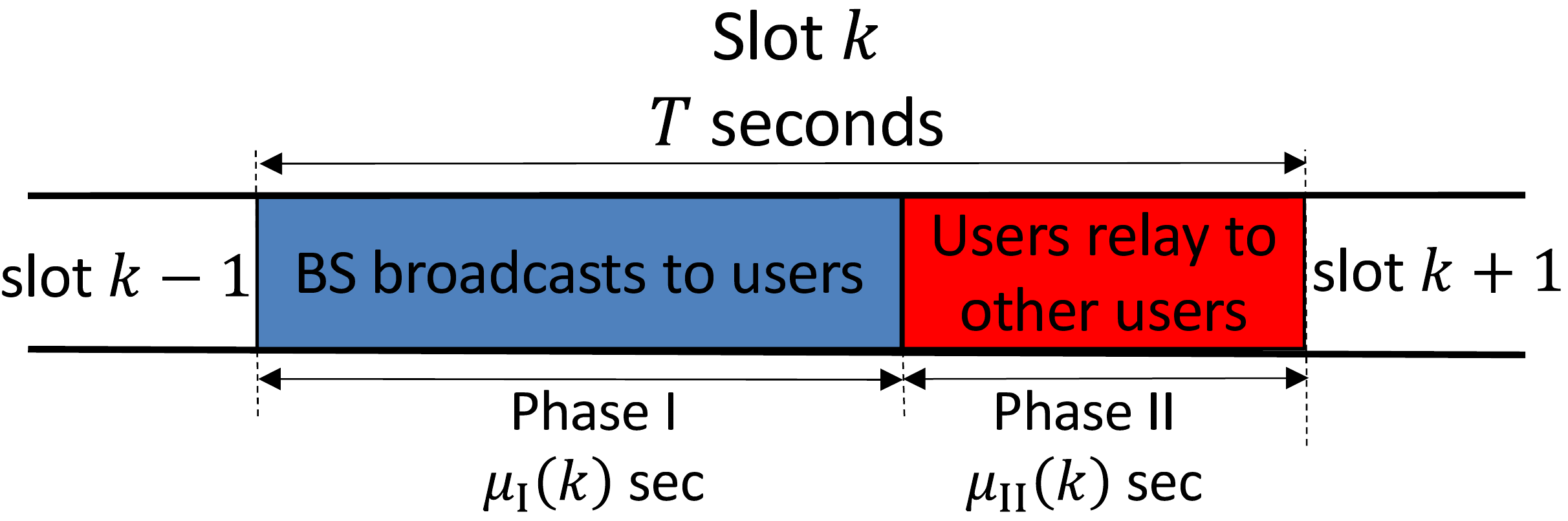}
\caption{Each time slot is divided into two phases; Phase I and Phase II. In Phase I, the BS broadcasts the packet to the users with a certain rate $\Rzk$. In Phase II, one of those users whose channel capacity was higher than $\Rzk$ and was able to decode in Phase I will be re-broadcasting this packet to give those who were not able to decode a second chance of potentially decoding the packet.}
\label{Time_Slot}%
\end{figure}

Following \cite{hou2010scheduling} we assume that more than one packet can be transmitted in one time slot. Thus, we divide time slot $k$ into two phases: Phase I and Phase II, with durations $\muRk$ and $\muNRk=\Ts-\muRk$, respectively (see Fig. \ref{Time_Slot}). In Phase I, the BS broadcasts the packets to its users with some rate $\Rzk$ given by
\begin{equation}
\Rzk=\log \lb 1+\PRzk\Gammazk\rb,
\label{Rate_z}
\end{equation}
where we normalize the noise variance of all receivers in the system to unity while $\Gammazk$ is referred to as the BS's ``gain threshold'' which is a parameter that is dictated by the BS's transmission rate $\Rzk$. Due to the fading nature of the channels, those users having their channel gains $\gammazik$ less than $\Gammazk$ are in outage and thus will not be able to decode this packet in Phase I. The smaller the rate $\Rzk$ is, the more users will be able to decode the packet in Phase I, but the more time it will take the BS to transmit the packet. In Phase II, one of these successful users, say user $\ist$, rebroadcasts the packet to potentially increase the number of users who decode it by the deadline. The transmission rate in Phase II by user $\ist$ is given by
\begin{equation}
R_{\ist}(k)=\log \lb 1+\PRistk\Gammaistk\rb.
\label{Rate_i}
\end{equation}
where $\Gammaistk$ is Phase II's ``gain threshold'' that is dictated by user $\ist$'s rate. Users with gain $\gammaistjk$ greater than $\Gammaistk$ will be able to decode the packet in Phase II. This technique offloads the data from the BS since it allows the users to help each other using Device-to-Device (D2D) communication while freeing up the BS during Phase II to serve other group of users outside the set $\sN$. Our objective in this paper is to maximize the long-term average duration of Phase II.

\section{Problem Formulation}
\label{Problem_Formulation}
\subsection{Objective Function}
At the beginning of the $k$th slot, the BS needs to decide the transmission rate $\Rzk$ by which it transmits in Phase I, the duration $\muRk$, the user $\ist$ that will relay the packet in Phase II as well as its transmission rate $R_{\ist}(k)$. This is what we refer to as the ``offloading decision problem'' which needs to be solved at the beginning of each time slot. The objective of this problem is to maximize the ``offloading factor'' which is the time-average value of $\muNRk$ and is given by
\begin{equation}
\bmuNR\triangleq\liminf_{K\rightarrow \infty}\frac{1}{K}\sum_{k=1}^K\frac{\muNRk}{\Ts}.
\label{Off_Fac}
\end{equation}
This value represents the average portion of the time slot that the BS is able to free by offloading the data to the local network of users. This portion of the time slot can be used to serve other users or increase the system capacity by adding more users.

\subsection{Constraints}
In addition to maximizing the BS's offloading factor, the users should be given a minimum QoS in terms of the average number of packets each was able to successfully decode by the deadline. We define $\bmuRi$ to be the average number of packets that user $i$ successfully decoded by the deadline and is given by
\begin{equation}
\bmuRi\triangleq \liminf_{K\rightarrow \infty}\sum_{k=1}^K\frac{\Oneik}{K}.
\label{Rate_RT}
\end{equation}
where
\begin{equation}
\Oneik=\left\{
\begin{array}{lll}
	&1 & \gammazik\geq\Gammazk \mbox{ OR } \gamma_{\ist i}(k)\geq \Gammaistk\\
	&0 & \mbox{otherwise}
\end{array}
\right.
\label{mu_ist}
\end{equation}
 is the indicator function which is $1$ if user $i$ was able to successfully decode the packet in Phase I or Phase II of slot $k$, and $0$ otherwise. Thus, the mathematical problem becomes
\begin{align}
\label{Prob_Offload}
\text{maximize } &\bmuNR,\\
\text{subject to }&\bmuRi\geq\lambda_i q_i, \hspace{1cm}\forall i\in\sN,
\label{RT_QoS}\\
&\text{At most 1 user transmits in Phase II},
\label{One_Tx}\\
&\muRk+\muNRk=\Ts.
\label{Ph_I_II_Dur}
\end{align}
Constraint \eqref{RT_QoS} indicates that the average number of packets decoded by user $i$ by the deadline is greater than the required QoS $q_i$, while constraint \eqref{One_Tx} indicates that at most one user should be allowed to transmit in Phase II.

\subsection{Degrees of Freedom}
Since the packet length of each packet in the system is fixed to $L$ bits, the degrees of freedom in this problem are 2, namely $\muRk$ and $\ist$. The reason is because once we find the value of $\muRk$ the BS's transmission rate $\Rzk$ can be calculated through the relation
\begin{equation}
\Rzk=\frac{L}{\muRk}.
\label{Rate_Duration_Eq}
\end{equation}
Similarly, once the user $\ist$ has been decided, the rate $R_{\ist}(k)$ can be found through the relation
\begin{equation}
R_{\ist}(k)=\frac{L}{\Ts-\muRk}.
\label{Rate_ist_Duration_Eq}
\end{equation}
Hence, the offloading problem in \eqref{Prob_Offload} constitutes of two coupled subproblems; the rate allocation problem of finding $\muRk$ as well as the scheduling problem of finding $\ist$. knowing the channel gains $\gammaijk$ with $i\in\{0\}\cup\sN$ and $j\in\sN$ in a negligible duration, the BS decides the duration of $\muNRk$ as well as the user $\ist$ that will be broadcasting the packet in Phase II. as well as $\muNRk$ through \eqref{Ph_I_II_Dur}.

We are interested in finding the (slot-based) rate allocation algorithm that maximizes the ``\emph{offloading factor}'' which is the average value of $\muNRk$, subject to the system constraints.

\section{Proposed Solution}
\subsection{Approach}
\label{Approach}

We propose to solve this problem using Lyapunov optimization \cite{li2011delay,Ewaisha_TVT2017}. We do this on three steps: i) We define a ``virtual queue'' associated with each average constraint in problem \eqref{Prob_Offload}. This helps in decoupling the problem across time slots. ii) Then we define a Lyapunov function, its drift and a, per-slot, reward function. iii) Based on the virtual queues and the Lyapunov function, we form and solve an optimization problem, for each slot $k$, that minimizes the drift-minus-reward expression. The solution of this problem is the proposed algorithm. We mathematically show the optimality of this algorithm.

We define the virtual queues $\Yik$ and $Z(k)$ as
\begin{align}
Y(k+1)&\triangleq \lb \Yik+a(k)q_i - \Oneik\rb^+,
\label{Yik}\\
Z(k+1)&\triangleq \lb Z(k)+r(k)-\muNRk\rb^+
\label{Zk}.
\end{align}
where $r(k)$ is an auxiliary variable that is to be optimized over. Its range is in the interval $[0,1]$. The queue $\Yik$ is an indication of how much user $i$ has been served from slot $1$ up to slot $k$. The larger the virtual queue $\Yik$ is, the more indication that user $i$ has not been served enough up to slot $k-1$, the more priority user should be given in slot $k$. On the other hand, $Z(k)$ indicates whether we should give priority to maximizing the offloading factor or to serving the users, during slot $k$.

To provide a sufficient condition on the virtual queues to satisfy the corresponding constraints, we use the definition of \emph{mean rate stability} of queues \cite[Definition 1]{li2011delay} to state the following lemma.

\begin{lma}
\label{Mean_Rate_Lemma}
If, for some $i\in\sN$, $\{\Yik\}_{k=0}^\infty$ is mean rate stable, then constraint \eqref{RT_QoS} is satisfied for user $i$.
\end{lma}
Lemma \ref{Mean_Rate_Lemma} shows that when the virtual queue $\Yik$ is mean rate stable, then constraint \eqref{RT_QoS} is satisfied for user $i\in\sN$. Similarly, if $\{Z(k)\}_{k=0}^\infty$ is mean rate stable, then we have
\begin{equation}
\liminf_{K\rightarrow\infty}\sum_{k=1}^K\frac{r(k)}{K}\leq\liminf_{K\rightarrow\infty}\sum_{k=1}^K\frac{\muNRk}{K}.
\label{Z_MRS}
\end{equation}
In the proof of the optimality of the proposed algorithm, we will see that \eqref{Z_MRS} is one of the keys to show this optimality. Thus, our objective now would be to devise an algorithm that guarantees the mean rate stability of both $\left[Y_i(k)\right]_{i\in\sN}$ and $Z(k)$.

\subsection{Applying the Lyapunov Optimization}
\label{Motivation_DL}
Let the quadratic Lyapunov function be defined as
\begin{equation}
L_{\rm yap}\lb U(k)\rb\triangleq \frac{1}{2}\sum_{i\in\sN}{Y_i^2(k)}+\frac{1}{2}{Z^2(k)},
\label{Lyapunov_Func}
\end{equation}
where $\bfU(k)\triangleq [\bfYk,Z(k)]$, and the Lyapunov drift as $\Delta (k) \triangleq \E_{U(k)}[L_{k+1}\lb {\bf U}(k+1)\rb - L_{\rm yap}\lb \bfU(k)\rb]$ where $\EEU{x}\triangleq \EE{x\vert U(k)}$ is the conditional expectation of the random variable $x$ given $U(k)$. Squaring \eqref{Yik} and \eqref{Zk}, taking the conditional expectation then summing over $i$, the drift becomes bounded by
\begin{align}
\nonumber\Delta(k)\leq & C+\sum_{i\in\sN}\EEU{\lb\Yik a(k)q_i - \Yik \Oneik\rb}\\
&+\lb\EEU{Z(k)r(k) - Z(k)\muNRk}\rb,
\label{Drift_Bound}
\end{align}
where $C\triangleq\lb\sum_{i\in\sN}\lb q_i^2+1\rb+1+\Ts^2\rb/2$. We then define $V$ as an arbitrarily chosen positive control parameter that controls the performance of the algorithm. Since $\EEU{r(k)}$ represents the expected duration of $\muNRk$ at slot $k$, we refer to $V \EEU{r(k)}$ as the ``reward term''. We subtract this term from both sides of \eqref{Drift_Bound}, then use \eqref{Psi_k} and rearrange to bound the drift-minus-reward term as
\begin{equation}
\Delta(k)-V \sum_{i\in\sN}\EEU{r(k)}\leq C+\Psi(k),
\label{Drift_minus_Reward_Bound}
\end{equation}
where
\begin{align}
\nonumber\Psi(k)\triangleq &-\EEU{\sum_{i\in\sN}\Yik\Oneik + Z(k)\muNRk}\\
&+\sum_{i\in\sN}\Yik \lambda q_i+\EEU{\lb Z(k)-V\rb r(k)}.
\label{Psi_k}
\end{align}
The algorithm we propose is to allocate the transmission rate and schedule the users to minimize the right-hand-side of \eqref{Drift_minus_Reward_Bound} at each slot. Since the only term in $\Psi(k)$ that is a function in $r(k)$ is the last term, we can decouple the problem without losing optimality. Minimizing this term results in setting $r(k)=1$ if $Z(k)<V$ and $0$ otherwise. Minimizing the remaining terms yields
\begin{equation}
\begin{array}{ll}
	&\text{maximize}\sum_{i\in\sN}\Yik\Oneik + Z(k)\muNRk\\
	&\text{subject to } \eqref{Ph_I_II_Dur} \text{ and } \eqref{One_Tx},
\end{array}
\label{Max_Prob}
\end{equation}
with decision variables $\muRk$ and $\ist$. This is a per-slot optimization problem the solution of which is an algorithm that minimizes the upper bound on the drift-minus-reward term defined in \eqref{Drift_minus_Reward_Bound}. Next we present the proposed algorithm.

\subsection{``\emph{Free-Base-Station}'' Algorithm}
The proposed algorithm to problem \eqref{Prob_Offload} is:
\begin{algorithm}
\caption{Free-BS Algorithm}
\begin{algorithmic}[1]
\label{Scheduling_Alg_Cont}
\STATE At the beginning of slot $k$, sort the users in a descending order of $\gammazik$. Without loss of generality, we assume that $\gammazik>\gamma_{0j}(k)$ for $i<j$.
\STATE Set $i=1$.
\WHILE{$i\leq N$}
\STATE Set $\Gammazk=\gammazik$ and calculate $\Rzk$, $\muRk$ and $\muNRk$ using \eqref{Rate_z}, \eqref{Rate_Duration_Eq} and \eqref{Ph_I_II_Dur}, respectively.
\STATE For all $j\leq i$, assume $j$ rebroadcasts the packet at Phase II and calculate the corresponding $\tilde{\Psi}_j(i)\triangleq\sum_{i\in\sN}\Yik\Oneik + Z(k)\muNRk$.
\STATE Calculate $\tilde{\Psi}(i)\triangleq\min_j \tilde{\Psi}_j(i)$.
\ENDWHILE
\STATE The optimum $\muRk$ comes from the iteration $i$ solving $\max_i\tilde{\Psi}(i)$, and $\ist=\arg\min_j \tilde{\Psi}_j(i)$.
\STATE Update \eqref{Yik} and \eqref{Zk} at the end of the $k$th slot.
\end{algorithmic}
\end{algorithm}

We can see that the algorithm calculates $\tilde{\Psi}_j(i)$ at most $N^2$ times yielding a polynomial time complexity. The performance of this algorithm is discussed next.
\subsection{Optimality of the Proposed Algorithm}
\begin{thm}
\label{Optimality_Thm}
For any value $V>0$, there exists some finite constant $C$ such that the Free-Base-Station algorithm results in an offloading factor satisfying
\begin{equation}
\liminf_{K\rightarrow \infty}\sum_{k=1}^K\frac{\mu^*_{\rm II}(k)}{K} \geq  \muopt-\frac{C}{V},
\label{Optimality_Equ}
\end{equation}
where $\mu^*_{\rm II}(k)$ is the optimal value of $\muNRk$ solving \eqref{Max_Prob}, while $\muopt$ is the optimal objective function achieved by the optimal algorithm solving problem \eqref{Prob_Offload}. Moreover, the queues $\Yik$ and $Z(k)$ are mean-rate stable.
\end{thm}

\begin{proofsketch}
We show the proof sketch of \eqref{Optimality_Equ} and omit the queues' mean-rate stability proof due to lack of space. Equation \eqref{Optimality_Equ} is shown by considering an optimal genie-aided algorithm solving \eqref{Prob_Offload} and showing that, when applied to the problem, the corresponding $\Psioptk$ satisfies
\begin{equation}
\liminf_{K\rightarrow\infty}\frac{1}{K}\sum_{k=1}^K\EE{\Psioptk}\leq-V\muopt.
\label{Key_Opt_Alg}
\end{equation}
Dropping $\Delta(k)$ from \eqref{Drift_minus_Reward_Bound}, evaluating by the Free-BS algorithm, taking $\EE{\cdot}$ to both sides, summing over $k$ and taking the limit yields
\begin{equation}
\liminf_{k\rightarrow\infty}\frac{1}{K}\sum_{k=1}^K\EE{r(k)}\leq C+\liminf_{k\rightarrow\infty}\frac{1}{K}\sum_{k=1}^K\EE{\Psi^*(k)},
\end{equation}
where $\Psi^*(k)$ is the value of $\Psi(k)$ when evaluated at the Free-BS algorithm. But since the Free-BS algorithm minimizes $\Psi(k)$, then we must have $\Psi^*(k)\leq\Psioptk$. Thus we can use the latter inequality and \eqref{Key_Opt_Alg} to write
\begin{equation}
V\liminf_{k\rightarrow\infty}\frac{1}{K}\sum_{k=1}^K\EE{r(k)}\geq V\muopt-C.
\label{Bound1}
\end{equation}
Removing the $(\cdot)^+$ sign from \eqref{Zk}, taking $\EE{\cdot}$ to both sides, summing over $k$ and taking the limit yields
\begin{equation}
\liminf_{k\rightarrow\infty}\frac{1}{K}\sum_{k=1}^K\EE{r(k)}\leq\liminf_{k\rightarrow\infty}\frac{1}{K}\sum_{k=1}^K\EE{\mu^*_{\rm II}(k)}.
\label{Bound_Using_Zk}
\end{equation}
Using \eqref{Bound1} and \eqref{Bound_Using_Zk} we get \eqref{Optimality_Equ}.
\end{proofsketch}

Theorem \ref{Optimality_Thm} indicates that setting the control parameter $V$ to a sufficiently high value results in an asymptotically optimal algorithm.

\section{Simulation Results}
\begin{figure}%
\centering
\includegraphics[width=1\columnwidth]{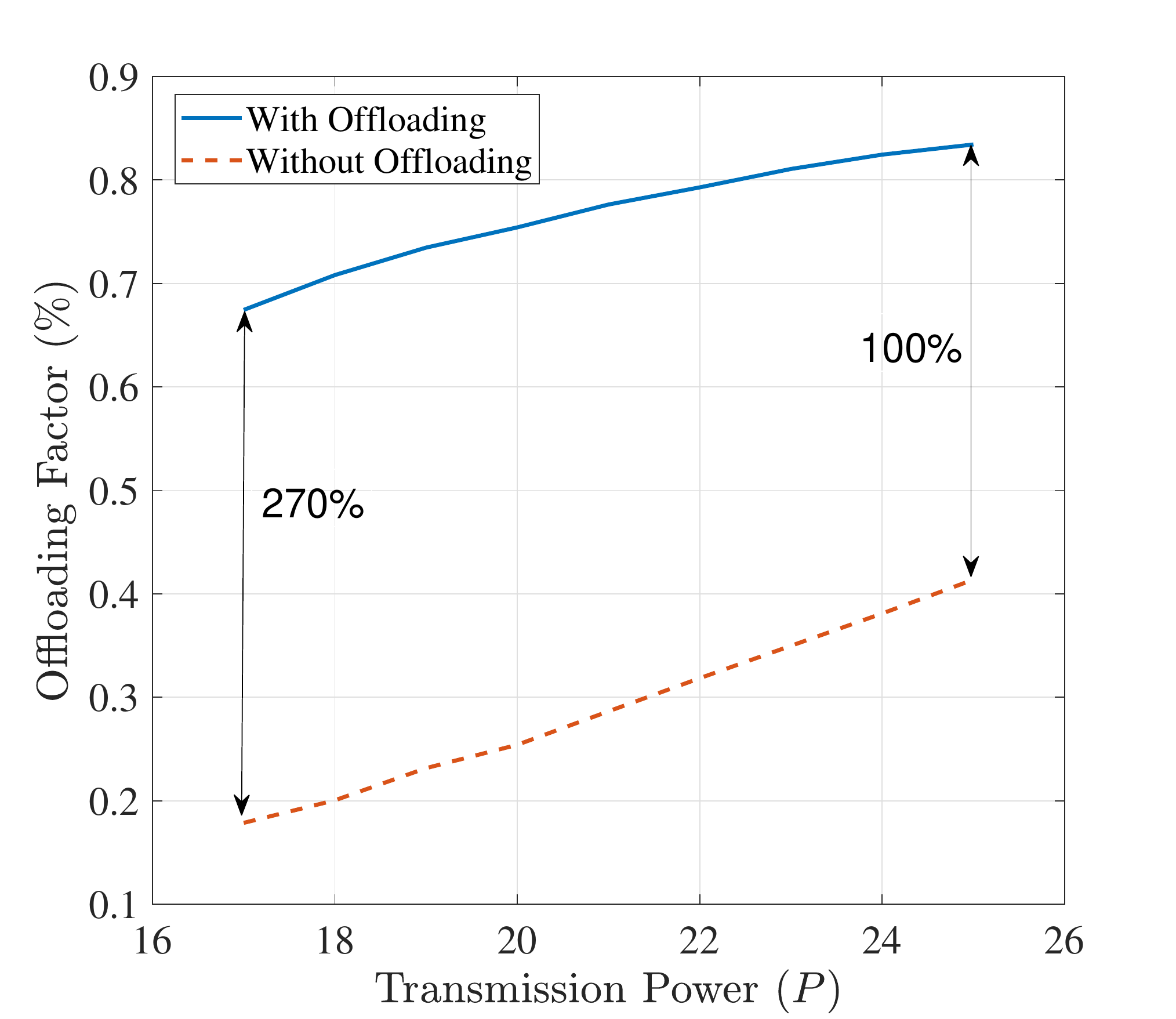}%
\caption{With just one user being allowed to relay, the BS is free $100\%$ times more.}%
\label{Offloading_vs_No_TxPow}%
\end{figure}

\begin{figure}%
\centering
\includegraphics[width=1\columnwidth]{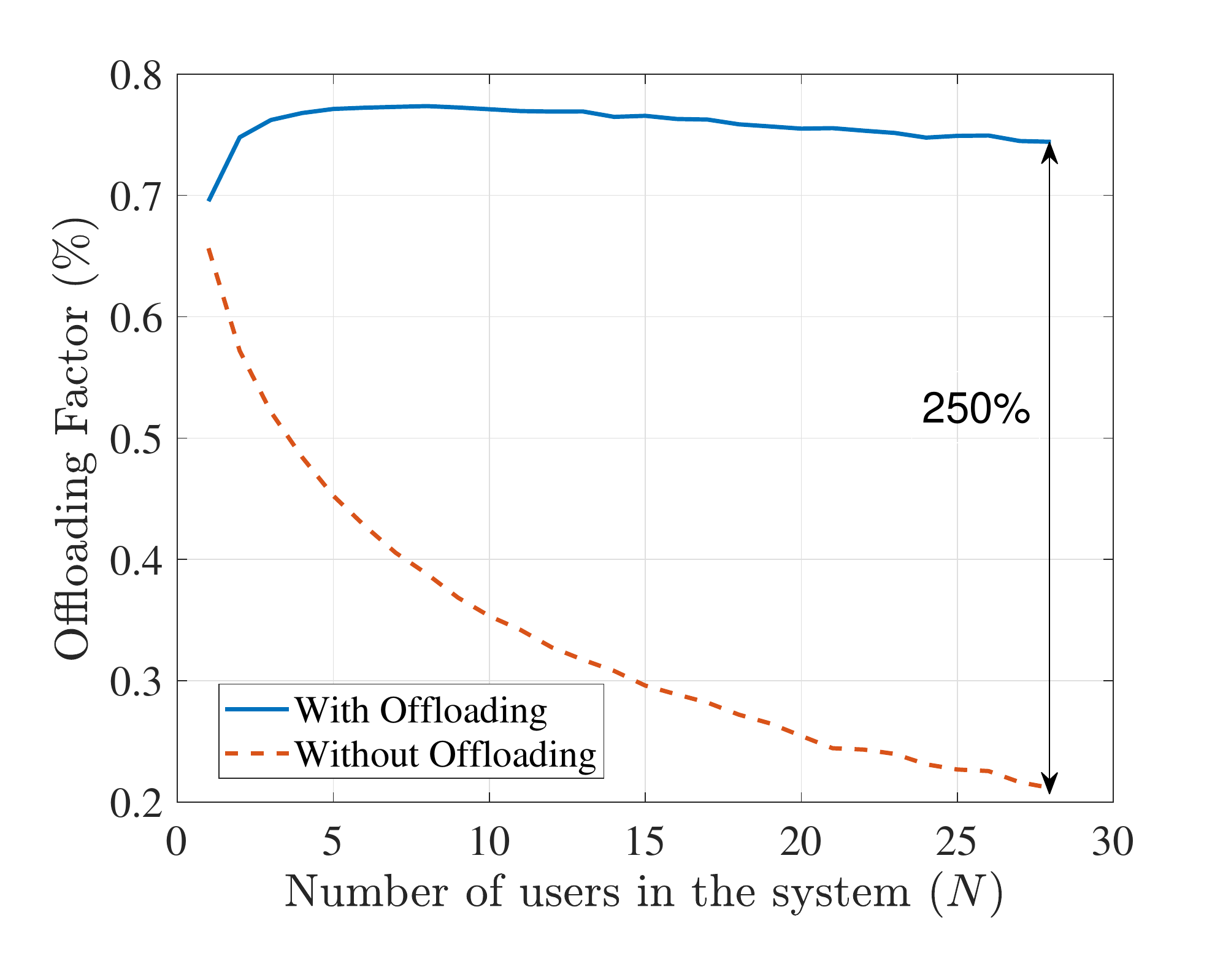}%
\caption{Unlike the ``Without Offloading'' case, when doing offloading, the throughput increases with $N$ for small values of $N$. This is due to the multi-user diversity effect where adding more users to the system creates more better users to retransmit the packet in Phase II of the time slot. This is a huge increase in the throughput with only a minor change in the system where only one user is allowed to retransmit.}%
\label{Offloading_vs_No_N}%
\end{figure}

The system was simulated with parameters shown in Table \ref{Parameters}. Fig. \ref{Offloading_vs_No_TxPow} shows the throughput performance with the transmission power. This figure shows that even when allowing for only one user to retransmit the packet in Phase II, the performance is significantly higher than the non-offloading case. In Fig. \ref{Offloading_vs_No_N} we plot the throughput versus the number of users in system ($N$). This figure shows that the non-offloading case has a decreasing throughput as $N$ increases. However, the offloading case is not monotonic under the proposed Free-BS algorithm. The throughput increases with $N$ when it the latter small due to the multi-user diversity effect \cite{zeng2012multi,6860297} where more users in the system gives the BS a larger set of users to choose from while scheduling Phase II's re-transmitting user. However, when $N$ increases beyond a certain value, adding more users to the system overloads it since these users need to be guaranteed a minimum average number of packets. Hence, the offloading factor starts decreasing.

\begin{table}
	\centering
		\caption{Simulation Parameter Values}
		\label{Parameters}
		\begin{tabular}{|c|c||c|c|}
			\cline{1-4}
			Parameter & Value & Parameter & Value \\
			\cline{1-4}
			$q_i$ & $0.9$ &$\Ts$ & $1$ms\\
			$L$ & $1$ bit/packet & $P_i$ $\forall i\in\{0\}\cup\script{N}$& $20$dB \\
			$V$ & $1000$ &$\bgammaij$ $\forall i,j$ & $0.3$
			\\ \cline{1-4}
			\end{tabular}
\end{table}

\section{Conclusions}
\label{Conclusions}
We discussed the problem of data offloading in cellular wireless systems. While existing work focuses on algorithms that offload the data locally to minimize traffic requested by cellular users, the objective of this work is to study the problem while taking the physical channel variations into consideration as well as the hard deadlines that have to be respected for each packet. We presented the Free-Base-Station algorithm to the formulated problem. In the full version we will show that it converges to the optimal solution asymptotically.


%



\bibliographystyle{IEEEbib}
\bibliography{MyLib}

\end{document}